\documentclass{jfm}

\usepackage{graphicx}
\usepackage{newtxtext}
\usepackage{newtxmath}
\usepackage{natbib}
\usepackage{hyperref}
\hypersetup{
    colorlinks = true,
    urlcolor   = blue,
    citecolor  = black,
}

\newcommand{\RomanNumeralCaps}[1]

\usepackage{hyperref} 
\usepackage[usenames, dvipsnames]{xcolor}
\hypersetup{breaklinks, colorlinks=true, linkcolor=blue, urlcolor=blue, citecolor=RoyalPurple}

\usepackage[font={footnotesize}]{caption}
\usepackage{bm}

\usepackage{scalerel, stackengine}
\setstackEOL{\#}
\stackMath
\def\hatgap{2pt}
\def\subdown{-2pt}
\newcommand\reallywidehat[2][]{ \renewcommand\stackalignment{l} \stackon[\hatgap]{#2}{ \stretchto{
    \scalerel*[\widthof{$#2$}]{\kern-.6pt\bigwedge\kern-.6pt}
    {\rule[-\textheight/2]{1ex}{\textheight}}}
    {0.5ex}_{\smash{ \belowbaseline[\subdown]{\scriptstyle#1} }}
}}

\renewcommand{\b}[1]    {\boldsymbol{#1}}
\renewcommand{\r}[1]    {\mathrm{#1}}

\renewcommand{\d}       {\partial}
\newcommand{\bu}        {\boldsymbol u}
\newcommand{\buS}       {\boldsymbol u^\mathrm{S}}
\newcommand{\uS}        {u^\mathrm{S}}

\newcommand{\bxh}       {\hspace{0.1em} \boldsymbol{\hat x}}
\newcommand{\byh}       {\hspace{0.1em}\boldsymbol{\hat y}}
\newcommand{\bzh}       {\hspace{0.1em}\boldsymbol{\hat z}}

\newcommand{\ee}        {\r{e}}

\newcommand{\dd}        {\r{d}}
\newcommand{\di}        {{\, \dd}}

\newcommand{\half}      {\tfrac{1}{2}}
\newcommand{\beq}       {\begin{equation}}
\newcommand{\eeq}       {\end{equation}}
\newcommand{\beqs}      {\begin{gather}}
\newcommand{\eeqs}      {\end{gather}}

\newcommand{\defn}      {\ensuremath{\stackrel{\r{def}}{=}}}

\renewcommand{\bnabla}    {\b{\nabla}}

\renewcommand{\bcdot}     {\b{\cdot}}

\newcommand{\bomega}     {\b{\omega}}
\newcommand{\bOmega}     {\b{\Omega}}

\newcommand{\com}       {\, ,}
\newcommand{\per}       {\, .}

\newcommand{\at}[1]     {\, |_{#1}}

\newcommand{\Ps}{\mathrm{Ps}}

\title{Phenomenology of decaying turbulence beneath surface waves}

\author{Gregory~L.~Wagner\aff{1}
  \corresp{\email{wagner.greg@gmail.com}}
 \and Navid~C.~Constantinou\aff{2, 3}}

\affiliation{\aff{1}Massachusetts Institute of Technology, Cambridge, MA, USA
\aff{2}University of Melbourne, Parkville, VIC, Australia
\aff{3}Australian Research Council Centre of Excellence for the Weather of the 21st Century, Australia}

\begin{document}
\maketitle

\begin{abstract}
This paper explores decaying turbulence beneath surface waves that is initially isotropic and shear-free.
We start by presenting phenomenology revealed by wave-averaged numerical simulations: an accumulation of angular momentum in coherent vortices perpendicular to the direction of wave propagation, suppression of kinetic energy dissipation, and the development of depth-alternating jets.
We interpret these features through an analogy with rotating turbulence \citep{holm1996ideal}, wherein the curl of the Stokes drift, $\bnabla \times \buS$, takes on the role of the background vorticity (for example, $(f_0 + \beta y) \bzh$ on the beta~plane).
We pursue this thread further by showing that a two-equation model  proposed by \cite{bardina1985effect} for rotating turbulence reproduces the simulated evolution of volume-integrated kinetic energy.
This success of the two-equation model --- which explicitly parametrizes wave-driven suppression of kinetic energy dissipation --- carries implications for modeling turbulent mixing in the ocean surface boundary layer.
We conclude with a discussion about a wave-averaged analogue of the Rossby number appearing in the two-equation model, which we term the ``pseudovorticity number'' after the pseudovorticity $\bnabla \times \bu^S$.
The pseudovorticity number is related to the Langmuir number in an integral sense.
\end{abstract}

\section{Introduction}

Surface waves enhance near-surface ocean turbulent mixing, especially in summertime and in the tropics where boundary layers are often shallow, sunny, windy, and wavy \citep{sullivan2010dynamics}.
Turbulence driven by surface wind stress and affected by surface waves is usually called ``Langmuir turbulence'' \citep{mcwilliams1997langmuir}, implying a connection between wave-catalyzed turbulent coherent structures and the structure of a laminar, wave-catalyzed shear instability that \cite{craik1976rational} called ``Langmuir circulation''.

In this paper we investigate decaying turbulence beneath surface waves using numerical simulations of the wave-averaged Navier--Stokes equations \citep{craik1976rational}.
Decaying turbulence is fundamental \citep{batchelor1953theory} but has seen little attention beneath surface waves.
Most work on wave-modified turbulence involves strong ambient shear and also invokes surface forcing by winds and buoyancy fluxes \citep{mcwilliams1997langmuir, polton2005role, harcourt2008large, van2012form, large2019similarity, fan2020effect}.
We show that some of the essential phenomenology of turbulence beneath surface waves is revealed by focusing on the evolution of unforced, initially-shear-free turbulence.

We interpret the results of these simulations in section~\ref{simulations}, leveraging an analogy between wave-averaged and rotating dynamics first proposed by \cite{holm1996ideal}.
Like rotation, surface waves \textit{catalyze} a transfer of energy from smaller to larger scales (hereby referred to as `inverse cascade') within the turbulent flow beneath the waves --- without exchanging energy with the flow --- and cause angular momentum to accumulate in coherent vortices, while suppressing kinetic energy dissipation.
This contrasts our case with the interaction between balanced flows and internal waves, wherein non-catalytic interactions can occur even for steady wave fields \citep{thomas2023turbulent}.
Moreover, unlike in shear-driven Langmuir turbulence where coherent structures tend to align with the direction of the Lagrangian-mean shear \citep{sullivan2010dynamics, van2012form}, the coherent vortices that we observe forming in the absence of shear are instead oriented \textit{perpendicular to the direction of surface wave propagation}.
We also observe the development of alternating jets, as in beta-plane turbulence.

In section~\ref{phenomenology} we further the analogy by adapting a phenomenological model proposed by \cite{bardina1985effect} for decaying rotating turbulence to the wave-averaged case, showing that the model also describes the decay of volume-averaged kinetic energy beneath surface waves.
This model suggests a new way to incorporate wave effects into two-equation models (for example, \cite{harcourt2015improved}) by explicitly representing the surface-wave-driven suppression of kinetic energy dissipation.

We conclude in section~\ref{discussion} by proposing a new non-dimensional number to characterize the importance of surface waves for the evolution of turbulent flows analogous to the Rossby number.
We show how this new number may be related to the Langmuir number for wind-forced cases, while also generalizing to purely convective or decaying situations.

\section{Simulations of decaying turbulence beneath surface waves}
\label{simulations}

The incompressible, inviscid momentum equation in the presence of a background vorticity~$\b{\Omega}$,
\beq
\d_t \bu + \left ( \bu \bcdot \bnabla \right ) \bu + \b{\Omega} \times \bu + \bnabla p = 0 \com \label{momentum}
\qquad \text{with} \qquad
\bnabla \bcdot \bu = 0 \com
\eeq
describes \emph{both} rotating flows without surface waves and \emph{also} wave-averaged flows beneath steady, horizontally-uniform surface-wave fields \citep[see for example][]{craik1976rational, suzuki2016understanding, holm1996ideal, wagner2021near}.
In the wave-averaged case,~$p$ and $\bu$ in~\eqref{momentum} represent the Eulerian-mean kinematic pressure and Lagrangian-mean velocity, respectively.
The Lagrangian-mean velocity is the ``total'' velocity responsible for the advection of mass, momentum, tracers, and particles, and may be decomposed into an Eulerian-mean component plus ``Stokes drift'' correction \citep[see for example]{van2018stokes, vanneste2022stokes}.
The wave-averaged momentum equation may be derived by asymptotic expansion leveraging a time-scale separation between rapidly oscillating surface waves and the slower evolution of $\bu$ \citep{craik1976rational, leibovich1980wave, vanneste2022stokes}.
The appearance of surface waves in~\eqref{momentum} as an effective Coriolis force is discussed by \citet{holm1996ideal}.

For rotating flows, the background vorticity $\b{\Omega}$ in~\eqref{momentum} takes the form
\beq
\b{\Omega}_\text{rotating} = f \bzh \com
\eeq
where $f$ is the Coriolis parameter and $\bzh$ is the axis of rotation.
For wave-averaged flows beneath a uniform surface wave field propagating in the $x$ direction, the background vorticity~$\b{\Omega}$ is alternatively
\beq
\b{\Omega}_\text{waves} = - \bnabla \times \buS = - \d_z \uS \byh \com
\eeq
where $\buS = \uS(z) \bxh$ is the surface wave Stokes drift \citep{craik1976rational, suzuki2016understanding, van2018stokes, vanneste2022stokes}.
We refer to $\bnabla \times \buS$ --- the curl of the surface-wave pseudomomentum $\buS$ \citep{andrews1978exact} --- as the ``pseudovorticity''.
In the context of equation~\eqref{momentum}, the main difference between rotating turbulence and turbulence beneath steady surface waves is the spatial structure of $\d_z \uS$: in shallow water $\d_z \uS \sim z$ to leading-order, while $\d_z \uS$ decays exponentially for deep water waves.


\subsection{Coherent structures in rotating and wave-averaged turbulence}

To illustrate the similarity of rotating turbulence and turbulence beneath surface waves we conduct large eddy simulations of equation~\eqref{momentum} with implicit grid-scale dissipation in a unit cube with three choices for $\bOmega$,
\beq \label{intro-stokes-shear}
\bOmega_\text{rotating} = \tfrac{1}{4} \bzh \com
\qquad
\bOmega_\text{waves} = -\tfrac{1}{2} z \byh \com
\quad \text{and} \quad
\bOmega_\text{isotropic} = 0 \per
\eeq
Above, $\tfrac{1}{2} z$ is the pseudovorticity associated with a shallow water wave in $z \in [0, 1]$ with Stokes drift $\buS \approx \tfrac{1}{2} \big (1 + \tfrac{1}{2} z^2 \big ) \bxh$.
The domain is horizontally-periodic in $x, y$, and we use free-slip boundary conditions at the top and bottom boundaries in $z$.
Implicit dissipation of grid-scale kinetic energy is provided by an upwind-biased, nominally 9th-order Weighted, Essentially Non-Oscillatory advection scheme \citep{pressel2017numerics, shu2020essentially, silvestri2024new}.
Our implicit dissipation method allows us to simulate decaying turbulence that undergoes a transition from isotropic to nearly two-dimensional dynamics.

\newcommand{\bk}{\boldsymbol{K}}

Following previous isotropic homogeneous turbulence studies (e.g., \citet{Orszag1972numerical}), we produce an initial condition for the three simulations by conducting a preliminary simulation initialized with kinetic energy spectrum
\beq
\half | \hat \bu |^2 \sim | \bk |^2 \exp \left \{-2 \left ( | \bk | / K_i \right )^2 \right \} \com
\eeq
where $\boldsymbol{\hat u}$ is the Fourier transform of $\bu$, $\boldsymbol{K}$ is the Fourier wavenumber vector, and $K_i = 32 \times 2\pi$ is the $32^\text{nd}$ wavenumber in the domain.
The amplitude of the preliminary initial condition is scaled to produce an initial root-mean-square vorticity
\beq \label{vorticity}
\sqrt{\int | \b{\omega} |^2 \di V} \defn \omega_\text{rms}(t=-t_0) = 1000 \com
\quad \text{where} \quad
\bomega \defn \bnabla \times \bu = \xi \bxh +  \eta \byh + \zeta \bzh \per
\eeq
The initial simulation is run for a duration $t_0$ until $t=0$, defined as the time when the mean-square vorticity has decayed to $\omega_\text{rms}(t=0) = 10$.
The velocity field is then saved to disk to be used as an initial condition in subsequent runs starting from $t=0$.
The simulations are conducted with Oceananigans \citep{ramadhan2020oceananigans, Wagner-etal-2025}, which discretizes~\eqref{momentum} with a finite volume method.
Scripts that reproduce simulations in this paper are stored on GitHub; see the Data availability statement.

\begin{figure}
    \centering
    \includegraphics[width = 1\textwidth]{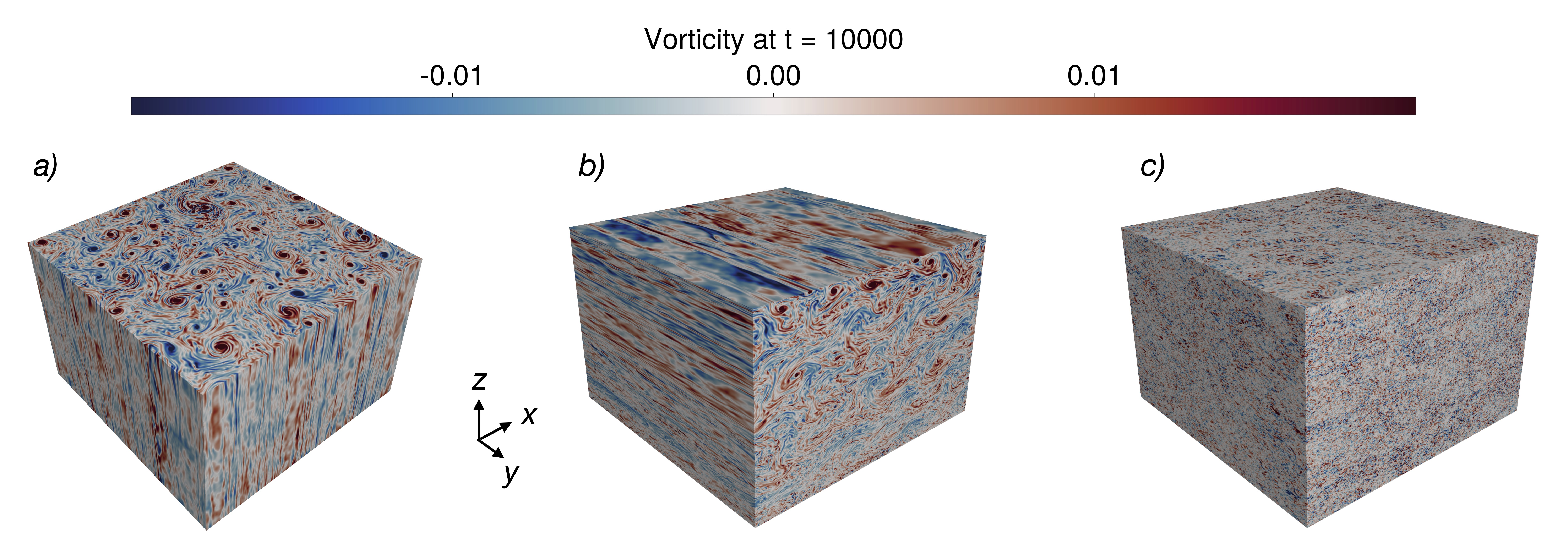}
    \caption{Vorticity in simulations decaying turbulence with $512^3$ finite volume cells in a unit cube after $1000$ time units.
    (a) Vertical relative vorticity $\zeta = \bzh \bcdot \left ( \bnabla \times \bu \right )$ in rotating turbulence with Coriolis parameter $f = 1/4$, (b) Horizontal relative vorticity $\eta = \byh \bcdot \left ( \bnabla  \times \bu \right )$ in turbulence beneath surface waves with Stokes shear $\d_z \uS = - z / 2$, and (c) $\eta$ in isotropic turbulence.}
    \label{decaying-turbulence}
\end{figure}

\begin{figure}
    \centering
    \includegraphics[width = 1\textwidth]{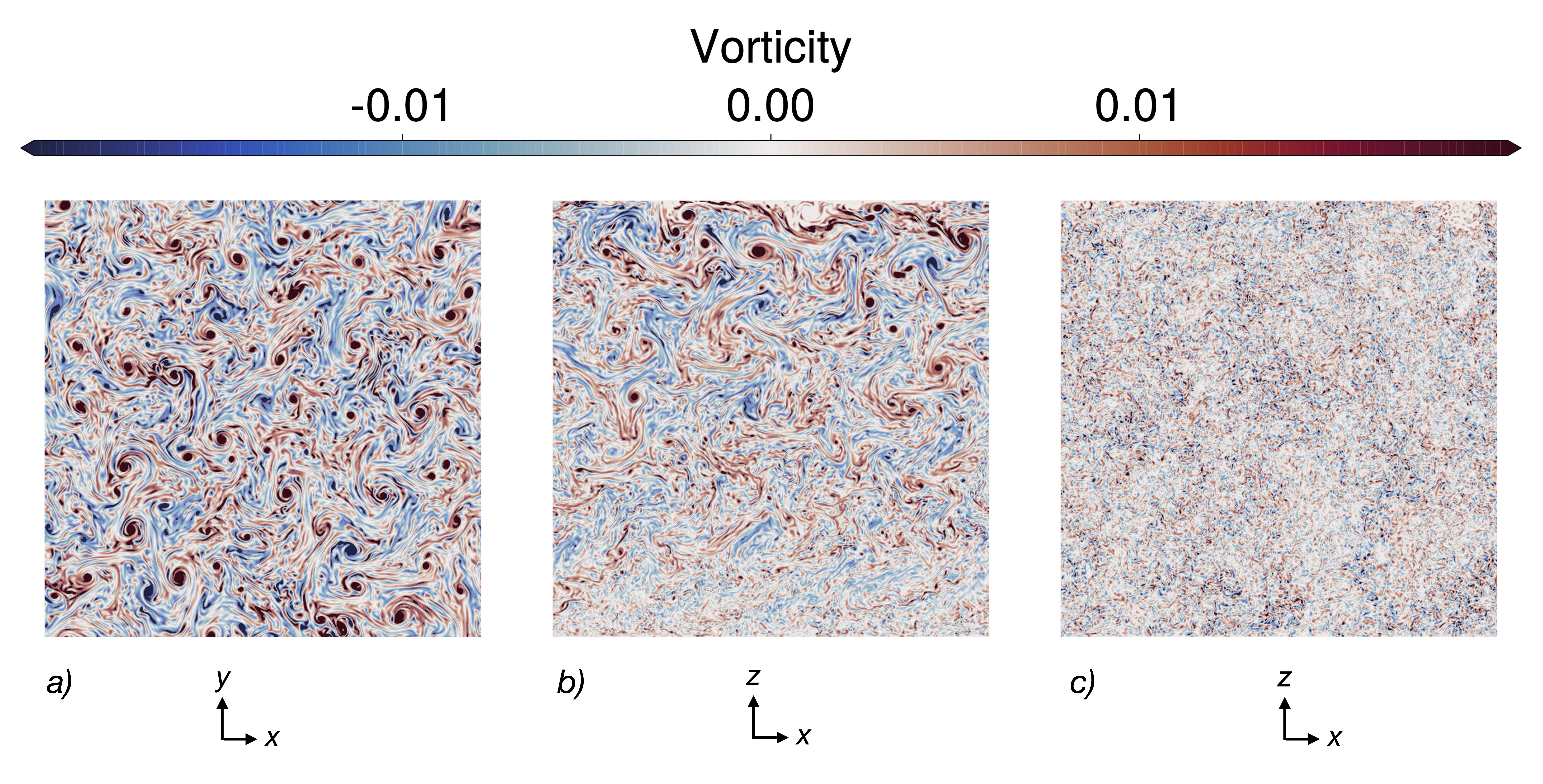}
    \caption{As figure~\ref{decaying-turbulence} but showing (a) $\zeta$ for rotating turbulence in the $xy$-plane; (b) $\eta$ for turbulence beneath surface waves in the $xz$-plane, (c) $\eta$ for isotropic turbulence in the $xz$-plane.}
    \label{decaying-turbulence-slice}
\end{figure}

We focus first on the evolution of the relative vorticity $\bomega$ defined in equation~\eqref{vorticity}.
Figure~\ref{decaying-turbulence} shows vorticity components for the three cases after $t=1000$ time units:
figure~\ref{decaying-turbulence}(a) shows vertical vorticity $\zeta$, while
while figures~\ref{decaying-turbulence}(b) and~(c) show the horizontal vorticity $\eta$.
Figure~\ref{decaying-turbulence}(a) and~(b) for rotating and wave-affected turbulence, respectively, both exhibit the formation of coherent structures and relatively greater vorticity levels than the unorganized, small amplitude isotropic vorticity in figure~\ref{decaying-turbulence}(c).
Figure~\ref{decaying-turbulence-slice} is similar, except that figure~\ref{decaying-turbulence-slice}(a) shows $\zeta$ in rotating turbulence in the $xy$-plane, while figures~\ref{decaying-turbulence-slice}(b) and~\ref{decaying-turbulence-slice}(c) show $\eta$ in the $xz$ plane for wave-averaged and isotropic turbulence, respectively.

\subsection{Zonation and the analogy of wave-averaged turbulence with beta-plane turbulence}

Note that turbulence beneath shallow water waves is homeomorphic to $\beta$-plane turbulence \cite{rhines1975waves} --- with $\bOmega_\text{$\beta$-plane} = \beta y \bzh$ --- modulo on the $xz$~plane instead of on the $xy$~plane and with the Stokes drift curvature $\d_z^2 \uS$ playing the role of $\beta$.
One of the most striking results of this similarity is the propensity for turbulence beneath surface waves to develop ``zonal jets'' --- coherent, alternating jets in the direction of surface wave propagation (and perpendicular to the pseudovorticity direction). This similarity allows us, therefore, to borrow intuition on structure formation in $\beta$-plane turbulence (for example, \cite{huang1998two, Farrell-Ioannou-2007, Srinivasan-Young-2012, Constantinou-etal-2014}; for an overview see \cite{Constantinou-2015, Farrell-Ioannou-2019, marston-tobias-2023}.)

To illustrate this, we consider two additional cases with background vorticity
\beq
\bOmega_\mathrm{deep} = - \tfrac{1}{4} \ee^{8 (z - 1)} \byh \per
\qquad \text{and} \qquad
\bOmega_\mathrm{weak} = - \tfrac{1}{8} z \byh \per
\eeq
These and subsequent simulations use $384^3$ finite volume cells to reduce their computational expense for the purpose of performing longer simulations to $t = 10^4$.

\begin{figure}
    \centering
    \includegraphics[width = 1\textwidth]{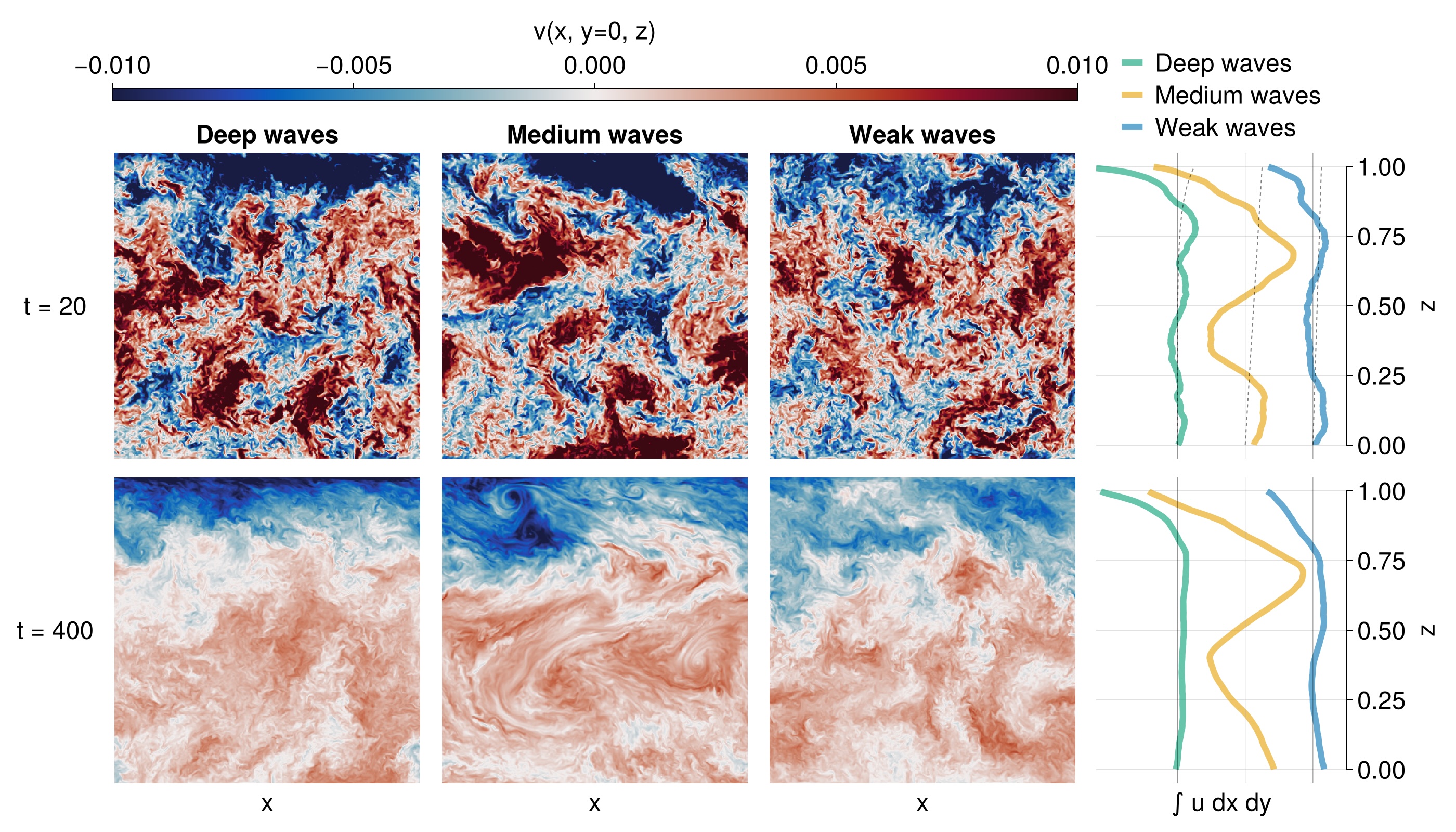}
    \caption{The evolution of cross-wave momentum $v$ in the $xz$-plane (6 panels on the left) and horizontally-averaged along-wave-momentum $u$ (2 panels on the right) at $t = 40, 400$ and for three wave fields: ``deep'' (left panel, red lines), ``medium'' (middle panel, orange lines), and ``weak'' (right panel, blue lines). Dashed lines in the top right plot show the Stokes drift profile (normalized) for each case. The ``medium'' waves case uses the background vorticity $\b{\Omega}_\text{waves}$ in~\eqref{intro-stokes-shear}. Both the $u$ profiles and the light gray ``zero lines'' are spaced apart by the scale $\delta u = 10^{-2}$. The $u$-profiles at right show the development of depth-alternating jets, including a counter-wave surface jet for all cases. The $v$-slices are similar between the three cases at early times, but at later times exhibit strong, localized wave-impacts in their respective regions of significant Stokes shear. Note that $v$ plays the role that vertical velocity plays in rotating turbulence.
    Note that the Rhines scale may be estimated as $\ell_R = 2 \pi \sqrt{U / \big | \d_z^2 \uS \big |}$. With $\big |\d_z^2 \uS \big | = \tfrac{1}{2}$ and $U \approx 10^{-2}$ for the medium waves case, for example, we estimate $\ell_R \approx 0.8$, consistent with the jet spacing apparent on the right panels.}
    \label{wave-averaged-evolution}
\end{figure}

Figure~\ref{wave-averaged-evolution} shows time-series of the $y$-momentum $v$, which plays the role that vertical velocity plays in rotating turbulence.
The right side of figure~\ref{wave-averaged-evolution} shows vertical profiles of the horizontally-averaged $x$-momentum,
\beq
U(z, t) \defn \int u \, \mathrm{d} x \, \mathrm{d} y \per
\eeq
These $U$-profiles exhibit the development of depth-alternating jets, which are most prominently exhibited by the zig-zagging structure in the medium waves case on the right side of figure~\ref{wave-averaged-evolution}.
To our knowledge, depth-alternating jets have not been observed in shear-free wave-modified turbulence.
Comparing the slices of $v$ for medium and weak waves reveals how the medium-strong waves induce a strong inverse cascade, more coherent vortices, and fewer small-scale motions than the weak waves.

\section{Phenomenological model for the evolution of kinetic energy}


\label{phenomenology}

\newcommand{\ddt} {\frac{\mathrm{d}}{\mathrm{d} t}}

\begin{figure}
    \centering
    \includegraphics[width = 0.8\textwidth]{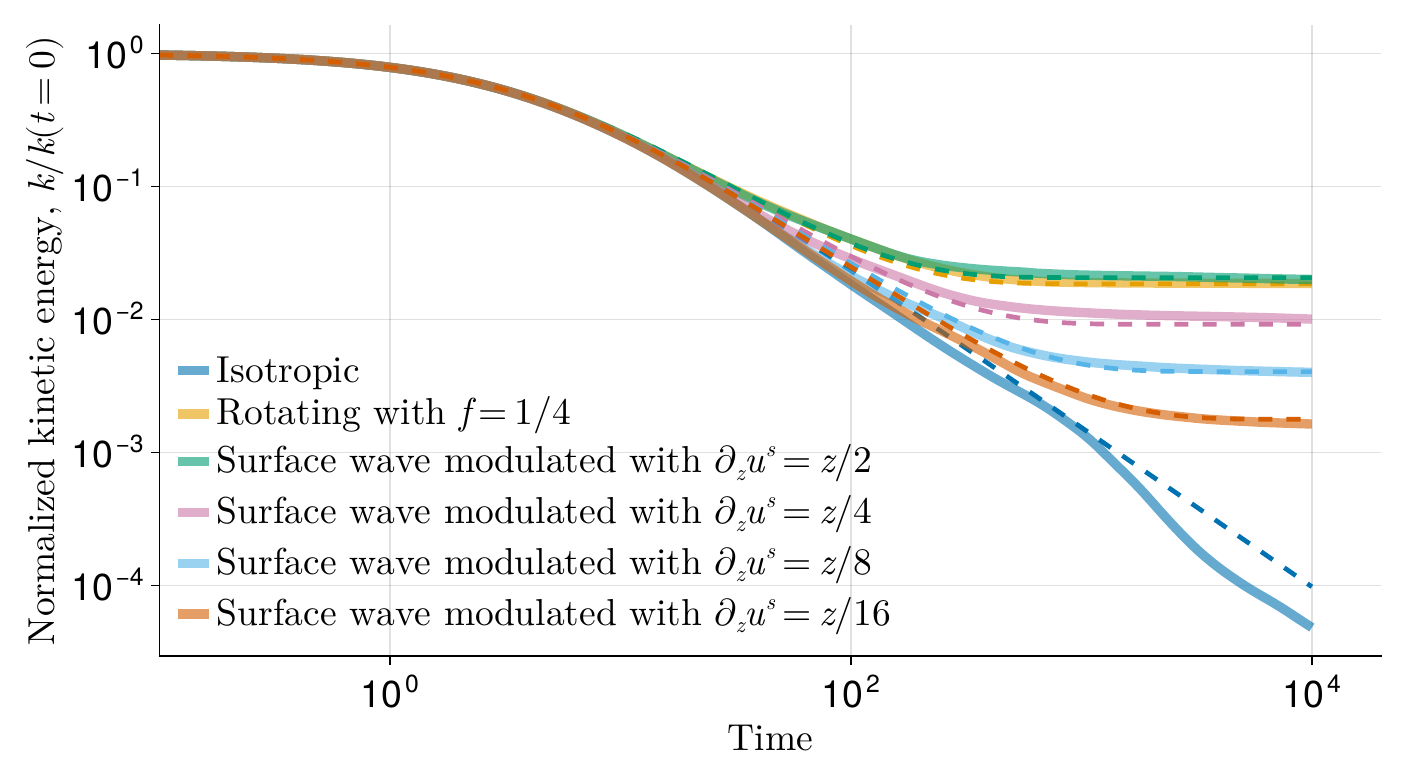}
    \caption{The decay of kinetic energy $k(t)$ in isotropic, rotating, and surface-wave-modulated turbulence. Solid lines show kinetic energy normalized by it's initial value, $k / k_0$, computed from large eddy simulations. Dashed lines show $k / k_0$ given by~\eqref{vorticity-solution}, which solves the phenomenological two-equation system in~\eqref{kinetic-energy-eqn}--\eqref{dissipation}. The initial dissipation rate $\epsilon_0$ in~\eqref{vorticity-solution} is estimated from the numerical solution; specifically we evaluate~\eqref{isotropic-solution} at $\Delta t$, substitute $k(t=\Delta t)$ from the numerical solution, and solve for $\epsilon_0$. We use $a = 11/6$ and estimate $b$ from~\eqref{beta-estimate}. For the rotating case, $\Omega = 1/4$ yields $b = 0.033$. For the four surface-wave-modulated cases we use $b = 0.036$, where $\Omega \defn \int \d_z \uS \di z = \left (1/4, 1/8, 1/16, 1/32 \right )$. Note that using $a = 1.75$ along with commensurate adjustments to $b$ matches the simulation data even more closely.}
    \label{kinetic-energy}
\end{figure}

We turn to the evolution of the domain-averaged kinetic energy,
\beq
k(t) \defn \int \half \left ( u^2 + v^2 + w^2 \right ) \di V \per
\eeq
Figure~\ref{kinetic-energy} plots time-series of the normalized kinetic energy $k(t) / k(t=0)$ for the three cases presented in figure~\ref{decaying-turbulence}, as well as three additional cases that use $\bOmega = S z \byh$ with $S = (1/2, 1/8, 1/16)$.
Figure~\ref{kinetic-energy} illustrates another common feature to rotating turbulence and turbulence beneath surface waves: a suppression of the kinetic energy dissipation rate, such that at long times the kinetic energy levels off.
We theorize that the suppression of dissipation is linked to the formation of coherent structures: rather than cascading to the grid scale, where dissipation can occur, kinetic energy accumulates in much larger-scale coherent structures.
Similar to the dynamics of two-dimensional turbulence \cite{mcwilliams1984emergence}, the concentration of kinetic energy in coherent structures suppresses the forward cascade of kinetic energy.

Inspired by \cite{bardina1985effect}, we model the kinetic energy $k(t)$ with the phenomenological two-equation system
\begin{align}
\ddt k &= -\epsilon \com \label{kinetic-energy-eqn} \\
\ddt \epsilon &= - a \frac{\epsilon^2}{k} - b \Omega \epsilon \com \label{dissipation}
\end{align}
where $\epsilon(t)$ describes the dissipation rate of $k$, $\Omega$ is a characteristic scale for $\b{\Omega}$, and $a$ and $b$ are $O(1)$ free parameters.
Equation~\eqref{kinetic-energy-eqn} follows directly from $\int \bu \bcdot \eqref{momentum} \di V$ after accounting for the effect of the implicit numerical dissipation in our simulations.
(Alternative models that include surface-wave-associated source terms for turbulent kinetic energy, such as the one proposed by \cite{axell2002wind}, are inconsistent with~\eqref{momentum}.)
The catalytic nature of rotation or surface waves may also be deduced from the fact that $\b{\Omega}$ does not appear in equation~\eqref{kinetic-energy-eqn}.
The first term in~\eqref{dissipation} models the destruction of $\epsilon$ on the turbulent time-scale $\tau = k / \epsilon$, since $\mathrm{d} \epsilon / \mathrm{d} t \sim - \epsilon / \tau $ when $\Omega \epsilon \ll \epsilon / \tau$.

The second term in \eqref{dissipation} models the suppression of kinetic energy dissipation --- or alternatively, the growth of the correlation length due to the coalescence of coherent structures --- by the background vorticity $\Omega$.
The relative importance of ``intrinsic'' destruction of $\epsilon$ and background-vorticity-induced destruction of $\epsilon$ is measured by the non-dimensional number $\Omega \tau = \Omega k / \epsilon$, whose significance is revisited in section~\ref{discussion}.
We note that \eqref{kinetic-energy-eqn} is exact --- and unchanged whether or not the system is rotating or modulated by surface waves.
The modulation of turbulence by surface waves is therefore fundamentally ``catalytic'', and can only affect $k$ indirectly by changing \eqref{dissipation}.

The free parameter $a$ may be constrained by considering  isotropic turbulence with $\Omega=0$.
In this case we expect the turbulent kinetic energy to decay according to $k \sim t^{-6/5}$ \citep{saffman1967large}, which implies $\ddt k = - \frac{6}{5} \frac{k}{t}$, and thus in turn, via \eqref{kinetic-energy-eqn}, leads to
\beq \label{isotropic-dissipation}
\epsilon = \frac{6}{5} \frac{k}{t} \com
\quad \text{so that} \quad
\ddt \epsilon = -\frac{66}{25} \frac{k}{t^2} \per
\eeq
Inserting~\eqref{isotropic-dissipation} into \eqref{dissipation} yields $a = 11/6$.



From \eqref{kinetic-energy-eqn}--\eqref{dissipation} we deduce that the rate of change of the combination $\log\epsilon - a \log k$ is proportional to $b$, that is,
\begin{gather} \label{intermediate-result}
\ddt \left ( \log \epsilon - a \log k \right ) = - b \Omega \per
\end{gather}
Integrating \eqref{intermediate-result} produces
\beq \label{vorticity-relation}
\frac{\epsilon}{\epsilon_0} =\left ( \frac{k}{k_0} \right )^a \ee^{- b \Omega t } \com
\eeq
where $\epsilon_0$ and $k_0$ are the dissipation and kinetic energy at $t = 0$ respectively.
If we insert the expression for the dissipation $\epsilon(t)$ from~\eqref{vorticity-relation} into~\eqref{kinetic-energy-eqn} and integrate in time, we obtain
\beq \label{vorticity-solution}
\frac{k}{k_0} = \left [ 1 + \frac{1}{n b \Omega} \frac{\epsilon_0}{k_0} \left ( 1 - \ee^{-b \Omega t} \right ) \right ]^{-n} \com
\eeq
where $n = 1 / (a - 1) = 6/5$.
(See also equation~(10) in \citet{bardina1985effect}.)
Taking the limit $\Omega \to 0$ (or $t \to 0$ with finite $\Omega$), we obtain the corresponding solution for isotropic turbulence,
\beq \label{isotropic-solution}
k_\text{isotropic}(t) = k_0 \left ( 1 + \frac{1}{n} \frac{\epsilon_0}{k_0} t \right )^{-n} \com
\eeq
which yields the expected power law $k \sim t^{-n}$ when $\epsilon_0 t/ n k_0 \gg 1$.

At long times the isotropic and vortical solutions diverge: $k_\text{isotropic} \to 0$ as $t \to \infty$, while in the vortical case $k/k_\infty$ limits to the constant
\beq \label{steady-state-tke}
\frac{k_\infty}{k_0} = \left ( 1 + \frac{\epsilon_0}{n b \Omega k_0} \right )^{-n} \com
\eeq
where $k_\infty = \lim_{t \to \infty} k(t)$.
Equation~\eqref{steady-state-tke} yields a formula for $b$ in terms of $k_\infty$,
\beq \label{beta-estimate}
b \Omega = \frac{\epsilon_0}{n k_0} \frac{1}{\left ( \frac{k_\infty}{k_0} \right )^{1/n} - 1} \com
\eeq
which we use to diagnose $b$ from our numerical simulations.

The dashed curves in figure~\ref{kinetic-energy} show solutions to the two-equation system in \eqref{kinetic-energy-eqn}--\eqref{dissipation} that correspond to the solid-line simulated results.
Figure~\ref{kinetic-energy} shows that a single value of $b = 0.036$ qualitatively describes the evolution of kinetic energy beneath surface waves for a wide range of background vorticity magnitudes $\Omega$, where we estimate $\Omega$ via the $z$-average of~$\d_z \uS$, that is $\Omega \approx \int \d_z \uS \di z$.
For the rotating case, we use $b = 0.033$ and $\Omega = 1/4$.
While qualitatively excellent considering that $b \approx 0.036$ describes a wide range of conditions, we also find that the phenomenological model overestimates the dissipation rate --- and therefore underestimates the kinetic energy~$k$ --- during the transition between isotropic and background-vorticity-dominated regimes.

\section{Discussion}
\label{discussion}

In this paper we point out the similarity between rotating turbulence on the beta-plane and turbulence beneath surface waves.
In particular, turbulence beneath surface waves exhibits the formation of coherent structures perpendicular to the direction of wave propagation, the development of zonal jets, and the suppression of kinetic energy dissipation.
These features are consistent with known properties of turbulence beneath surface waves, but the connection with rotating turbulence is obscured in studies that also involve surface wind stress, ambient Lagrangian-mean shear, and surface forcing.
In particular, we hypothesize that some of the generic features of Langmuir turbulence --- such as the formation of coherent structures, a suppression of kinetic energy dissipation, and an increase in mixing --- are driven by similar dynamics as the phenomenon observed in our decaying situations.
At the same time, however, the presence of Lagrangian-mean shear produces additional and distinctly different phenomenon, most clearly in the alignment of coherent vortices with the Lagrangian-mean shear, rather than perpendicular to the direction of wave propagation.

We exploit the connection to rotating turbulence by adapting a phenomenological two-equation model proposed by \cite{bardina1985effect}.
In this two-equation model, the evolution of dissipation is affected by two terms: one ``classical'' term producing power-law decay of kinetic energy, and a second term that describes the suppression of kinetic energy dissipation by the presence of surface waves, which in turn effectively enhances kinetic energy levels and turbulent mixing relative to pure isotropic turbulence.

This phenomenological model hints at a new way to understand how surface waves enhance turbulent mixing.
In the paradigm proposed by \cite{mcwilliams1997langmuir}, the effect of surface waves on turbulence is associated with the ``Stokes contribution'' to shear production in the turbulent kinetic energy budget.
However, our results show that this interpretation must be incomplete, because surface waves also control the evolution of initially shear-free flows.

We therefore propose an alternative theory which supposes that the impact of surface waves may be instead linked to their tendency to catalyze, without exchanging energy, an increase in the correlation times and length scales of turbulent motions.
A benefit to our alternative interpretation is that turbulent shear production can be interpreted in the standard way: as a transfer of kinetic energy from the horizontally-averaged, Lagrangian-mean velocity, whose energy is otherwise conserved.
In other words, our descriptive analysis leads to an alternative paradigm for wave-modified turbulence wherein shear production is ``unchanged'' relative to wave-free turbulence (as in rotating turbulence).
Instead of modifying shear production directly, then, waves impact both mixing and kinetic energy by inducing an inverse cascade, increasing the turbulent mixing length, and suppressing kinetic energy dissipation.

\subsection{The ``pseudovorticity'' number}

The analogy with rotating turbulence inspires the definition of a new non-dimensional number for characterizing surface wave effects on turbulence.
In rotating turbulence, the Rossby number measures the relative magnitude of the relative vorticity $\bnabla \times \bu$ and the background vorticity, $f \bzh$, such that
\beq
\mathrm{Ro} \defn \frac{| \bnabla \times \bu |}{f} \sim \frac{U}{f L} \com
\eeq
where $U$ is a characteristic horizontal velocity scale and $L$ is a characteristic turbulent horizontal scale.
We propose the analogous ``pseudovorticity number'' for boundary layer turbulence,
\beq \label{stokes-shear-number}
\mathrm{Ps} \defn \frac{| \bnabla \times \bu |}{|\d_z \buS|} \sim \frac{W}{\Omega H}\com
\eeq
where $W$ is a turbulence velocity scale, $\Omega$ is the magnitude of the Stokes drift shear, and $H$ is a turbulence length scale.
In boundary layer turbulence, for example, $W$ and $H$ are most straightforwardly measured by the magnitude of the vertical velocity and the boundary layer depth.

The pseudovorticity number~$\Ps$ measures the average role of surface waves with Stokes shear magnitude $\Omega$ on turbulent motions with turbulent velocity scale $W$ and turbulent length scale $H$.
In the two-equation model in~\eqref{kinetic-energy-eqn}--\eqref{dissipation}, $\Ps \sim \epsilon / k \Omega$.
In our decaying scenarios, $\Ps$ eventually vanishes at $t \to \infty$ and dissipation is completely suppressed.

\subsection{Connection with the Langmuir number}

\newcommand{\La}{\mathrm{La}}

The pseudovorticity number is connected between the Langmuir number --- the usual non-dimensional quantity used to characterize the effect of waves on turbulence --- when considering an estimate of $\Ps$ integrated over deep boundary layers.
One definition of the Langmuir number is \citep{mcwilliams1997langmuir},
\beq
\La \defn \sqrt{\frac{u_\star}{\uS(z=0)}} \com
\eeq
where $u_\star$ is the friction velocity, which is the square root of the kinematic wind stress.
At face value, $\La$ does not apply to decaying cases, but we can amend this by interpreting~$u_\star$ more generally as a turbulent velocity scale.

To connect $\La$ and $\Ps$, we consider a ``bulk'' estimate of $\Ps$ over a wind-forced boundary layer of depth $H$, where the turbulent velocity scale is $W = u_\star$ and the turbulent length scale is $H$.
If $\uS(z=-H)$ is negligible, then estimating the pseudovorticity scale $\Omega$ as the average value over the boundary layer yields
\beq
\Omega \sim \frac{1}{H} \int_{-H}^0 \d_z \uS \di z \sim \frac{\uS(z=0)}{H} \com
\eeq
such that
\beq
\text{bulk Ps} = \frac{W}{\Omega H} \sim \frac{u_\star}{\uS(z=0)} \sim \mathrm{La^2} \per
\eeq
We thus find that $\La^2$ may be regarded as a bulk estimate of the more locally-applicable~$\Ps$, computed over the depth of the boundary layer, $H$.
The difference between the bulk estimate~$\La$ and the more specific estimate~$\Ps$ resolves a paradox associated with~$\La$ in that it depends on $\uS$, despite that only $\d_z \uS$ appears in \eqref{momentum}.

\backsection[Acknowledgements]{Without implying their endorsement, we would like to acknowledge fruitful discussions with Bruno Deremble, Petros Ioannou, and Bill Young. We would also like to thank the editor and the three anonymous reviewers for their constructive comments that greatly improved our paper. Part of this work was done during the Sean~R.~Haney Memorial Symposium 2025 at the Scripps Institution of Oceanography in La Jolla, California.}

\backsection[Funding]{
G.L.W.~was supported by the National Science Foundation, grant OCE-2342715.
N.C.C.~was supported by the Australian Research Council under the DECRA Fellowship~DE210100749, the Center of Excellence for the Weather of the 21st Century~CE230100012, and the Discovery Project~DP240101274.}

\backsection[Declaration of interests]{The authors report no conflict of interest.}

\backsection[Data availability statement]{Scripts to reproduce the simulations and analyses  in this of this study are openly available at the GitHub repository \href{https://github.com/glwagner/WaveAveragedDecayingTurbulence}{github.com/glwagner/WaveAveragedDecayingTurbulence}.

    Simulation output used to produced the figures in this paper will be uploaded in a Zenodo repository upon acceptance.}

\backsection[Author ORCIDs]{Gregory~L.~Wagner, \href{https://orcid.org/0000-0001-5317-2445}{0000-0001-5317-2445}; Navid~C.~Constantinou, \href{https://orcid.org/0000-0002-8149-4094}{0000-0002-8149-4094}.}



\begin{thebibliography}{36}
\expandafter\ifx\csname natexlab\endcsname\relax\def\natexlab#1{#1}\fi
\def\au#1{#1} \def\ed#1{#1} \def\yr#1{#1}\def\at#1{#1}\def\jt#1{\textit{#1}} \def\bt#1{#1}\def\bvol#1{\textbf{#1}} \def\vol#1{#1} \def\pg#1{#1} \def\publ#1{#1}\def\arxiv#1{#1}\def\org#1{#1}\def\st#1{\textit{#1}}

\bibitem[Andrews \& McIntyre(1978)]{andrews1978exact}
{\sc \au{Andrews, D.~G.} \& \au{McIntyre, M.~E.}} \yr{1978}  \at{An exact theory of nonlinear waves on a {Lagrangian}-mean flow}.  \jt{J. Fluid Mech.}  \bvol{89}~(4),  \pg{609--646}.

\bibitem[Axell(2002)]{axell2002wind}
{\sc \au{Axell, L.~B.}} \yr{2002}  \at{Wind-driven internal waves and {Langmuir} circulations in a numerical ocean model of the southern {Baltic Sea}}.  \jt{J. Geophys. Res. Oceans}  \bvol{107}~(C11),  \pg{25--1}.

\bibitem[Bardina {\em et~al.\/}(1985)Bardina, Ferziger \& Rogallo]{bardina1985effect}
{\sc \au{Bardina, J.}, \au{Ferziger, J.~H.} \& \au{Rogallo, R.~S.}} \yr{1985}  \at{Effect of rotation on isotropic turbulence: computation and modelling}.  \jt{J. Fluid Mech.}  \bvol{154},  \pg{321--336}.

\bibitem[Batchelor(1953)]{batchelor1953theory}
{\sc \au{Batchelor, G.~K.}} \yr{1953} {\em The theory of homogeneous turbulence\/}.  \publ{Cambridge University Press}.

\bibitem[van~den Bremer \& Breivik(2018)]{van2018stokes}
{\sc \au{van~den Bremer, T.~S.} \& \au{Breivik, {\O}.}} \yr{2018}  \at{Stokes drift}.  \jt{Philos. T. R. Soc. A}  \bvol{376}~(2111),  \pg{20170104}.

\bibitem[Constantinou(2015)]{Constantinou-2015}
{\sc \au{Constantinou, N.~C.}} \yr{2015}  \at{Formation of large-scale structures by turbulence in rotating planets}. PhD thesis, National and Kapodistrian University of Athens, Athens, Greece.

\bibitem[Constantinou {\em et~al.\/}(2014)Constantinou, Farrell \& Ioannou]{Constantinou-etal-2014}
{\sc \au{Constantinou, N.~C.}, \au{Farrell, B.~F.} \& \au{Ioannou, P.~J.}} \yr{2014}  \at{Emergence and equilibration of jets in beta-plane turbulence: {Applications of Stochastic Structural Stability Theory}}.  \jt{J. Atmos. Sci.}  \bvol{71}~(5),  \pg{1818--1842}.

\bibitem[Craik \& Leibovich(1976)]{craik1976rational}
{\sc \au{Craik, A. D.~D.} \& \au{Leibovich, S.}} \yr{1976}  \at{A rational model for {Langmuir} circulations}.  \jt{J. Fluid Mech.}  \bvol{73}~(3),  \pg{401--426}.

\bibitem[Fan {\em et~al.\/}(2020)Fan, Yu, Savelyev, Sullivan, Liang, Haack, Terrill, de~Paolo \& Shearman]{fan2020effect}
{\sc \au{Fan, Y.}, \au{Yu, Z.}, \au{Savelyev, I.}, \au{Sullivan, P.~P.}, \au{Liang, J.-H.}, \au{Haack, T.}, \au{Terrill, E.}, \au{de~Paolo, T.} \& \au{Shearman, K.}} \yr{2020}  \at{The effect of {Langmuir} turbulence under complex real oceanic and meteorological forcing}.  \jt{Ocean Model.}  \bvol{149},  \pg{101601}.

\bibitem[Farrell \& Ioannou(2007)]{Farrell-Ioannou-2007}
{\sc \au{Farrell, B.~F.} \& \au{Ioannou, P.~J.}} \yr{2007}  \at{Structure and spacing of jets in barotropic turbulence}.  \jt{J. Atmos. Sci.}  \bvol{64}~(10),  \pg{3652--3665}.

\bibitem[Farrell \& Ioannou(2019)]{Farrell-Ioannou-2019}
{\sc \au{Farrell, B.~F.} \& \au{Ioannou, P.~J.}} \yr{2019}  \at{{Statistical State Dynamics: A new perspective on turbulence in shear flow}}.  \bt{In {\em {Zonal jets: Phenomenology, genesis, and physics}\/} (ed. \ed{B.~Galperin \& P.~L. Read})},  \pg{p. 380–400}.  \publ{Cambridge University Press}.

\bibitem[Harcourt(2015)]{harcourt2015improved}
{\sc \au{Harcourt, R.~R.}} \yr{2015}  \at{An improved second-moment closure model of {Langmuir} turbulence}.  \jt{J. Phys. Oceanogr.}  \bvol{45}~(1),  \pg{84--103}.

\bibitem[Harcourt \& D’Asaro(2008)]{harcourt2008large}
{\sc \au{Harcourt, R.~R.} \& \au{D’Asaro, E.~A.}} \yr{2008}  \at{Large-eddy simulation of {Langmuir} turbulence in pure wind seas}.  \jt{J. Phys. Oceanogr.}  \bvol{38}~(7),  \pg{1542--1562}.

\bibitem[Holm(1996)]{holm1996ideal}
{\sc \au{Holm, D.~D.}} \yr{1996}  \at{The ideal {Craik--Leibovich} equations}.  \jt{Physica D}  \bvol{98}~(2-4),  \pg{415--441}.

\bibitem[Huang \& Robinson(1998)]{huang1998two}
{\sc \au{Huang, H.-P.} \& \au{Robinson, W.~A.}} \yr{1998}  \at{Two-dimensional turbulence and persistent zonal jets in a global barotropic model}.  \jt{J. Atmos. Sci.}  \bvol{55}~(4),  \pg{611--632}.

\bibitem[Large {\em et~al.\/}(2019)Large, Patton, DuVivier, Sullivan \& Romero]{large2019similarity}
{\sc \au{Large, W.~G.}, \au{Patton, E.~G.}, \au{DuVivier, A.~K.}, \au{Sullivan, P.~P.} \& \au{Romero, L.}} \yr{2019}  \at{Similarity theory in the surface layer of large-eddy simulations of the wind-, wave-, and buoyancy-forced southern ocean}.  \jt{J. Phys. Oceanogr.}  \bvol{49}~(8),  \pg{2165--2187}.

\bibitem[Leibovich(1980)]{leibovich1980wave}
{\sc \au{Leibovich, S.}} \yr{1980}  \at{On wave-current interaction theories of langmuir circulations}.  \jt{J. Fluid Mech.}  \bvol{99}~(4),  \pg{715--724}.

\bibitem[Marston \& Tobias(2023)]{marston-tobias-2023}
{\sc \au{Marston, J.~B.} \& \au{Tobias, S.~M.}} \yr{2023}  \at{Recent developments in theories of inhomogeneous and anisotropic turbulence}.  \jt{Annu. Rev. Fluid Mech.}  \bvol{55},  \pg{351--375}.

\bibitem[McWilliams(1984)]{mcwilliams1984emergence}
{\sc \au{McWilliams, J.~C.}} \yr{1984}  \at{The emergence of isolated coherent vortices in turbulent flow}.  \jt{J. Fluid Mech.}  \bvol{146},  \pg{21--43}.

\bibitem[McWilliams {\em et~al.\/}(1997)McWilliams, Sullivan \& Moeng]{mcwilliams1997langmuir}
{\sc \au{McWilliams, J.~C.}, \au{Sullivan, P.~P.} \& \au{Moeng, C.-H.}} \yr{1997}  \at{Langmuir turbulence in the ocean}.  \jt{J. Fluid Mech.}  \bvol{334},  \pg{1--30}.

\bibitem[Orszag \& Patterson(1972)]{Orszag1972numerical}
{\sc \au{Orszag, S.~A.} \& \au{Patterson, G.~S.}} \yr{1972}  \at{Numerical simulation of three-dimensional homogeneous isotropic turbulence}.  \jt{Phys. Rev. Lett.}  \bvol{28},  \pg{76--79}.

\bibitem[Polton {\em et~al.\/}(2005)Polton, Lewis \& Belcher]{polton2005role}
{\sc \au{Polton, J.~A.}, \au{Lewis, D.~M.} \& \au{Belcher, S.~E.}} \yr{2005}  \at{The role of wave-induced {Coriolis--Stokes} forcing on the wind-driven mixed layer}.  \jt{J. Phys. Oceanogr.}  \bvol{35}~(4),  \pg{444--457}.

\bibitem[Pressel {\em et~al.\/}(2017)Pressel, Mishra, Schneider, Kaul \& Tan]{pressel2017numerics}
{\sc \au{Pressel, K.~G.}, \au{Mishra, S.}, \au{Schneider, T.}, \au{Kaul, C.~M.} \& \au{Tan, Z.}} \yr{2017}  \at{Numerics and subgrid-scale modeling in large eddy simulations of stratocumulus clouds}.  \jt{J. Adv. Model. Earth Syst.}  \bvol{9}~(2),  \pg{1342--1365}.

\bibitem[Ramadhan {\em et~al.\/}(2020)Ramadhan, Wagner, Hill, Campin, Churavy, Besard, Souza, Edelman, Ferrari \& Marshall]{ramadhan2020oceananigans}
{\sc \au{Ramadhan, A.}, \au{Wagner, G.~L.}, \au{Hill, C.}, \au{Campin, J.-M.}, \au{Churavy, V.}, \au{Besard, T.}, \au{Souza, A.}, \au{Edelman, A.}, \au{Ferrari, R.} \& \au{Marshall, J.}} \yr{2020}  \at{{Oceananigans.jl: Fast and friendly geophysical fluid dynamics on GPUs}}.  \jt{J. Open Source Softw.}  \bvol{5}~(53),  \pg{2018}.

\bibitem[Rhines(1975)]{rhines1975waves}
{\sc \au{Rhines, P.~B.}} \yr{1975}  \at{Waves and turbulence on a beta-plane}.  \jt{J. Fluid Mech.}  \bvol{69}~(3),  \pg{417--443}.

\bibitem[Saffman(1967)]{saffman1967large}
{\sc \au{Saffman, P.~G.}} \yr{1967}  \at{The large-scale structure of homogeneous turbulence}.  \jt{J. Fluid Mech.}  \bvol{27}~(3),  \pg{581--593}.

\bibitem[Shu(2020)]{shu2020essentially}
{\sc \au{Shu, C.-W.}} \yr{2020}  \at{Essentially non-oscillatory and weighted essentially non-oscillatory schemes}.  \jt{Acta Numer.}  \bvol{29},  \pg{701--762}.

\bibitem[Silvestri {\em et~al.\/}(2024)Silvestri, Wagner, Campin, Constantinou, Hill, Souza \& Ferrari]{silvestri2024new}
{\sc \au{Silvestri, S.}, \au{Wagner, G.~L.}, \au{Campin, J.-M.}, \au{Constantinou, N.~C.}, \au{Hill, C.~N.}, \au{Souza, A.} \& \au{Ferrari, R.}} \yr{2024}  \at{A new {WENO}-based momentum advection scheme for simulations of ocean mesoscale turbulence}.  \jt{J. Adv. Model. Earth Syst.}  \bvol{16}~(7),  \pg{e2023MS004130}.

\bibitem[Srinivasan \& Young(2012)]{Srinivasan-Young-2012}
{\sc \au{Srinivasan, K.} \& \au{Young, W.~R.}} \yr{2012}  \at{Zonostrophic instability}.  \jt{J. Atmos. Sci.}  \bvol{69}~(5),  \pg{1633--1656}.

\bibitem[Sullivan \& McWilliams(2010)]{sullivan2010dynamics}
{\sc \au{Sullivan, P.~P.} \& \au{McWilliams, J.~C.}} \yr{2010}  \at{Dynamics of winds and currents coupled to surface waves}.  \jt{Annu. Rev. Fluid Mech.}  \bvol{42}~(1),  \pg{19--42}.

\bibitem[Suzuki \& Fox-Kemper(2016)]{suzuki2016understanding}
{\sc \au{Suzuki, N.} \& \au{Fox-Kemper, B.}} \yr{2016}  \at{Understanding {Stokes} forces in the wave-averaged equations}.  \jt{J. Geophys. Res. Oceans}  \bvol{121}~(5),  \pg{3579--3596}.

\bibitem[Thomas(2023)]{thomas2023turbulent}
{\sc \au{Thomas, J.}} \yr{2023}  \at{Turbulent wave-balance exchanges in the ocean}.  \jt{Proc. R. Soc. A}  \bvol{479}~(2276),  \pg{20220565}.

\bibitem[Van~Roekel {\em et~al.\/}(2012)Van~Roekel, Fox-Kemper, Sullivan, Hamlington \& Haney]{van2012form}
{\sc \au{Van~Roekel, L.~P.}, \au{Fox-Kemper, B.}, \au{Sullivan, P.~P.}, \au{Hamlington, P.~E.} \& \au{Haney, S.~R.}} \yr{2012}  \at{{The form and orientation of Langmuir cells for misaligned winds and waves}}.  \jt{J. Geophys. Res. Oceans}  \bvol{117}~(C5).

\bibitem[Vanneste \& Young(2022)]{vanneste2022stokes}
{\sc \au{Vanneste, J.} \& \au{Young, W.~R.}} \yr{2022}  \at{Stokes drift and its discontents}.  \jt{Philos. T. R. Soc. A}  \bvol{380}~(2225),  \pg{20210032}.

\bibitem[Wagner {\em et~al.\/}(2021)Wagner, Chini, Ramadhan, Gallet \& Ferrari]{wagner2021near}
{\sc \au{Wagner, G.~L.}, \au{Chini, G.~P.}, \au{Ramadhan, A.}, \au{Gallet, B.} \& \au{Ferrari, R.}} \yr{2021}  \at{Near-inertial waves and turbulence driven by the growth of swell}.  \jt{J. Phys. Oceanogr.}  \bvol{51}~(5),  \pg{1337--1351}.

\bibitem[Wagner {\em et~al.\/}(2025)Wagner, Silvestri, Constantinou, Ramadhan, Campin, Hill, Chor, Strong-Wright, Lee, Poulin, Souza, Burns, Marshall \& Ferrari]{Wagner-etal-2025}
{\sc \au{Wagner, G.~L.}, \au{Silvestri, S.}, \au{Constantinou, N.~C.}, \au{Ramadhan, A.}, \au{Campin, J.-M.}, \au{Hill, C.}, \au{Chor, T.}, \au{Strong-Wright, J.}, \au{Lee, X.~K.}, \au{Poulin, F.}, \au{Souza, A.}, \au{Burns, K.~J.}, \au{Marshall, J.} \& \au{Ferrari, R.}} \yr{2025}  \at{{High-level, high-resolution ocean modeling at all scales with Oceananigans}}.  \jt{arXiv preprint} ,  \arxiv{arXiv: 2502.14148}.

\end{thebibliography}
\end{document}